\documentclass[11pt,twocolumn]{article}
\usepackage{graphicx,wrapfig}  
\begin{document}

\newcommand{\al}{\ensuremath{\alpha}}
\newcommand{\alc}{\ensuremath{\alpha_c}}
 
\hyphenation{Con-nec-ti-vi-ty E-vol-ving To-po-lo-gy 
na-tur-vid-en-ska-be-li-ge}

\title{Random Geometric Graphs}
\author{
Jesper Dall\thanks{Email: j.dall@fysik.sdu.dk} \ 
and Michael Christensen 
\\
Fysisk Institut, SDU--Odense Universitet\\
Campusvej 55, DK--5230 Odense M\\
Denmark
}
\date{\today}
\maketitle
 
\begin{abstract} 
\noindent  We analyse graphs in which each vertex
is assigned random coordinates in a geometric space
of arbitrary dimensionality and 
only edges between adjacent points are present.
The critical connectivity is found numerically
by examining the size of the largest cluster.
We derive an analytical expression 
for the cluster coefficient which shows
that the graphs are distinctly different
from standard random graphs, even for infinite dimensionality. 
Insights relevant for graph
bi-partitioning are included.
\end{abstract}

\section{Introduction}
The interest in complex networks has exploded over the last
f\mbox{}ive years~\cite{barabasi01,strogatz01} where data on very large
networks like 
the WWW~\cite{huberman99,newman01b,barabasi99},
collaborations in the scientif\mbox{}ic community~\cite{newman01c},
transportation~\cite{amaral00},
movie actor collaborations~\cite{watts98}
etc.\ have become accessible.

Random graphs are often used
to model complex networks~\cite{kauffman95}. 
Ever since Erd{\"o}s and R{\'e}nyi's groundbraking work more than forty
years ago~\cite{erdos60}, 
intense theoretical research on random graphs has been taking place
~\cite{newman01b,newman01d,bollobas85,kirkpatrick94}.
In contrast to random graphs the interactions between the sites
in a lattice are usually between nearest neighbours, reflecting a
myopic world. Lattices are therefore often said to be at the other 
end of the spectrum of network models~\cite{kleinberg00,stauffer92}.

Properties of real networks
like robustness~\cite{cohen00,albert00},
growth~\cite{newman01d,krapivsky00,albertbarabasi00,dorogovtsev00},
and topology have attracted much attention,
primarily from physicists.  
It has been consistently shown that many of the
networks possess small world
characteristics~\cite{watts98,watts99,newman99}.
Like random graphs, small world networks are characterized by
short average distances between any two sites, and by a high
degree of localness, much like in lattices.  However,
individually, random graphs and lattice models in their pure forms
are poor models of many real world networks.
One could argue that high-dimensional lattices have the
necessary high clustering and low average path
length, though this has not been explored much~\cite{newman00c}.
In the current paper we provide
results on high-dimensional systems.

A random geometric graph (RGG) is a random graph with a metric.
It is constructed by assigning each vertex random coordinates in
a $d$-dimensional box of volume 1, i.e.\ each coordinate 
is drawn from a uniform distribution on the unit interval. 
RGGs have been used sporadically in real networks 
modeling~\cite{banavar99} and extensively in continuum 
percolation~\cite{xia88,balberg85,
alonbalberg90,alonbalberg91,quintanilla00}, but almost exclusively
in two and three dimensions.
Although RGGs are the continuum version of lattices,
they deserve some attention of their own, since percolating continuum
systems display behaviour that lattices are incapable of~\cite{balberg87}.
In addition, the connectivity in RGGs can be 
increased in a more natural  way than by adding new bonds randomly
in lattices.

Recently, continuum percolation has been used in the
study of the stretched exponential decay of the correlation
function in random walks on fractals
and 
the conjectured relation to relaxation in complex systems~\cite{jund01}. 
However, continuous systems in general and RGGs in
particular are relevant whenever we need a multi-dimensional
system with a metric, as for example when modeling the spread of 
diseases~\cite{vespignani01c}.

In this paper we study RGGs in arbitrary dimensions. 
In low dimensions the systems are dominated by local interactions.
For higher dimensions RGGs are usually believed to approach 
standard random graphs, which we show is true only in some respects.
We focus on `phase transitions'~\cite{kirkpatrick94,cheeseman91,anderson99}
at the percolation threshold by looking at the size of the largest
cluster, and we determine how the value 
of the critical parameter in RGGs approaches that
of random graphs as the dimension increases. We also 
extract the distribution of cluster sizes in the critical region.
Furthermore, an expression for the cluster
coef\mbox{}f\mbox{}icient, a quantity that has 
attracted much interest in network theory recently, is derived.
Results relevant for graph bi-partitioning are established.
Finally, we discuss how to implement random geometric 
graphs ef\mbox{}f\mbox{}iciently. 

The layout of this paper is as follows. In Section~\ref{sec:RG} 
and \ref{sec:RGG} we describe random graphs
and random geometric graphs, respectively. In Section~\ref{sec:Results} we
present our results, and Section~\ref{sec:Implementation} contains
the details regarding the implementation. Finally, 
in Section~\ref{sec:Conclusion} we sum up.

\section{Random Graphs} \label{sec:RG}
\begin{figure}[t]
\begin{center}
\includegraphics[width=7.5cm]{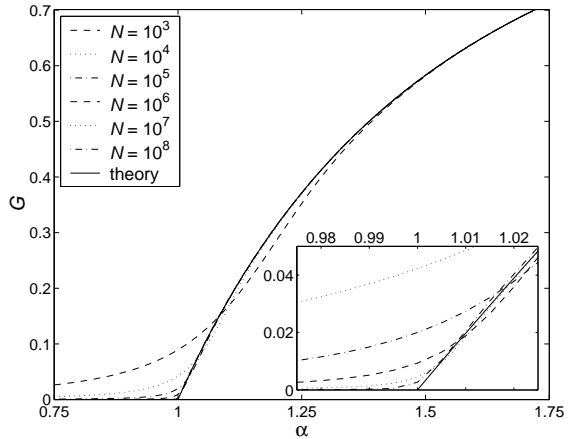}
\caption{\small 
	The size of the largest cluster in random graphs
	as a function of the connectivity.
	Note that for $N>10^6$ the Monte Carlo data is almost
	indistinguishable from the theoretical result in 
	Eq.~(\ref{alpha(G)}). Error bars  are not shown
	since they are in all cases less than the width of the lines.
	Inset: A closer look at the percolation threshold 
	$\alpha_c=1$.
	}
\label{RGlargestc}
\end{center}
\vspace{-0.5cm}
\end{figure}
Random graphs consist of $N$ vertices (points/sites) and $K$ 
edges (lines) where each possible edge is present with
probability $p$, i.e.\ $K=pN(N-1)/2$.\footnote{From here  
on we consider $N \simeq N-1$ in accordance
with the literature~\cite{newman01b,wong87}, since we are
only investigating large systems.}
To keep the discussion 
independent of the system size $N$, graphs are often characterized
by the connectivity (degree) $\alpha=2K/N=pN$, i.e.\ the average 
number of connections per vertex, instead of $K$ or $p$.
As the
connectivity increases clusters of vertices appear, where a cluster
consists of all vertices 
linked together by edges, directly or 
indirectly.

The size of the largest cluster in the macroscopic limit
\mbox{$N\rightarrow\infty$} can be calculated 
analytically~\cite{erdos60,bollobas85}. It is $NG(\alpha)$, where
\begin{equation}
	G(\alpha)=1-\frac{1}{\alpha}\sum_{n=1}^{\infty}
	\frac{n^{n-1}}{n!}(\alpha e^{-\alpha})^n.
\label{G(alpha)}
\end{equation}
By the use of~\cite{bollobas85}
\begin{equation}
	y=\sum_{n=1}^{\infty}\frac{n^{n-1}}{n!}x^n \ \  
	\Leftrightarrow
	\ \ x=ye^{y}
\label{helpRG}
\end{equation}
we can invert Eq.~(\ref{G(alpha)}), getting
\begin{equation}
	\alpha(G)=-\frac{1}{G}\log(1-G).
\label{alpha(G)}
\end{equation}
from which it is trivial to show that $\alpha_c = 1$.
With Eq.~(\ref{alpha(G)}) 
it is an easy task to plot the fraction of vertices in 
the largest cluster---the
giant component---as done
in Fig.~\ref{RGlargestc}, where we see the prototype of a phase 
transition in combinatorial problems.

In random graphs the probability distribution 
of 
edges $p_k$ is binomial
\begin{equation}
	p_k =\left(\!\! 
	\begin{array}{c}
		N \\ k
	\end{array}
	\!\!\right) p^{k}(1-p)^{N-k} \simeq
	 \frac{\alpha^k e^{-\alpha}}{k!},
\label{poisson}
\end{equation}
\noindent where the approximation resulting
in the Poisson distribution is valid
for large systems sizes $N$, which is exactly
the limit in which we are interested. 
The critical connectivity $\alpha_c$ for graphs with arbitrary 
random degree distribution $p_k$ has recently been derived by other
techniques than those originally leading to
Eq.~(\ref{G(alpha)})~\cite{newman01b,molloyreed95,molloyreed98}.
Unfortunately, we cannot use these results in connection with 
random geometric graphs, 
as will become clear in the next section.

\section{Random Geometric Graphs} \label{sec:RGG}
\begin{figure}[t]
\begin{center}
\includegraphics[width=7.5cm]{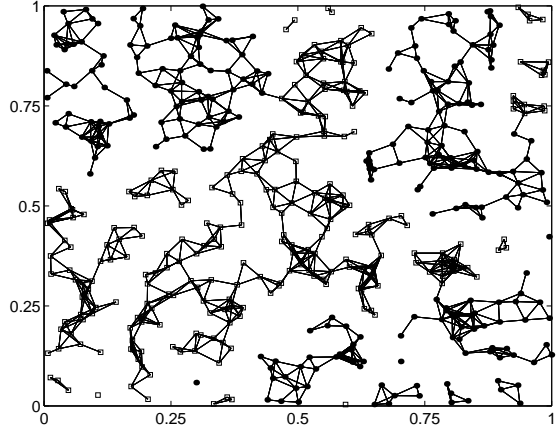}
\caption{\small 
	A $2D$ random geometric graph with  $N=500$ and $\alpha=5$. The
	graph is bi-partitioned---see Section~\ref{sec:GBP}. 
	There are no edges across the boundaries, i.e.\ the
	boundary conditions are open, not continuous.
	}
\label{ggplot5}
\end{center}
\vspace{-0.5cm}
\end{figure}
A $d$-dimensional random geometric graph (RGG) is a graph where each
of the $N$ vertices is assigned random coordinates in the box
$[0,1]^d$, and only points `close' 
to each other are connected by an edge. The degree distribution
of a RGG with average connectivity $\alpha$ is therefore given
by  Eq.~(\ref{poisson}) as well.
However, a RGG is a special kind of random graph with properties 
not captured by the theoretical tools mentioned above.
For one thing, 
the probability that three vertices are cyclically connected
is dif\mbox{}ferent in random graphs and RGGs, regardless of 
the degree distribution of the random graph.

RGGs are sometimes named spatial graphs~\cite{watts98}.
Fig.~\ref{ggplot5} illustrates a RGG in $2D$.
As in lattices, dif\mbox{}ferent boundary conditions can be applied.
We will see that toroidal (continuous) 
boundary conditions make a vital dif\mbox{}ference
compared to having open boundary conditions.

The volume of a $d$-dimensional (hyper)sphere with radius 
$r$ is
\begin{equation}
V_{sphere}=\frac{{\pi}^{d/2}r^{d}}{\Gamma(\frac{d+2}{2})},
\label{volume_d}
\end{equation}
\noindent where $\Gamma(x)$ is the gamma function.
This volume is needed in order to f\mbox{}ind the edges in RGGs. 
\begin{figure}[t]
\begin{center}
\includegraphics[width=7.5cm]{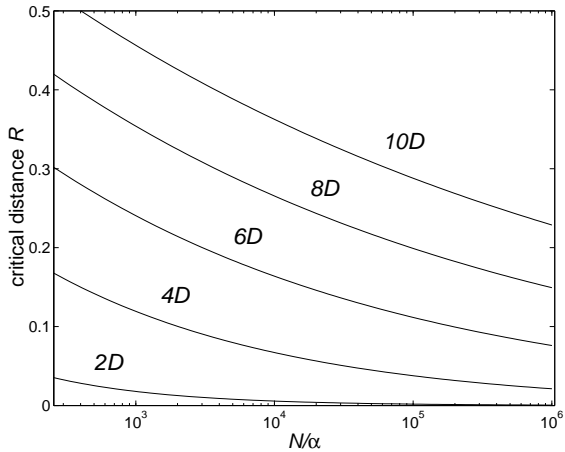}
\caption{\small The critical distance in random geometric graphs 
		in various dimensions.
		Points within this distance of each other are
		connected by an edge. The critical distance is
		equivalent to the radius $R$ of the excluded
		volume associated with each point. 
		}
 \label{critradius}
\end{center}
\vspace{-0.5cm}
\end{figure}

To `visualise' a RGG in general, 
one can think of a box f\mbox{}illed with small spheres
with  radius $r$ and volume $V$ given by Eq.~(\ref{volume_d}), 
where points are connected by an edge only if the distance
between their centers is $<2r$, i.e.\ if the spheres overlap. 
Since the total volume of our box is 1, 
the probability that two arbitrarily chosen vertices 
are connected is equal to the volume of a sphere with
radius $R=2r$. In continuum percolation theory this volume
is denoted the \emph{excluded volume} $V_{ex}$, where
$V_{ex}=2^dV$ in a RGG.
The excluded volume is the basic quantity of interest because
it is directly related to the connectivity
\begin{equation}
	\alpha=Np=NV_{ex},
\label{alfa&volume_ex}
\end{equation}
from which it is clear why the connectivity is
frequently called the total excluded volume of the system.
Eqs.~(\ref{volume_d}) and~(\ref{alfa&volume_ex}) give us
\begin{equation}
	R=\frac{1}{\sqrt{\pi}}
	\left[ 
	\frac{\alpha}{N}\Gamma\left( \frac{d+2}{2} \right) 
	\right]^{\frac{1}{d}}.
\label{alfatoradius}
\end{equation}
Fig.~\ref{critradius} shows the radius $R$ 
of the excluded volume
as a function of $N/\alpha=1/p=1/V_{ex}$.
 $R$ decreases monotonically: for a given
connectivity $\alpha$ the spheres have to become smaller
when more vertices are added to the graph.

Eq.~(\ref{alfatoradius}) provides us with the required 
relation  between  $\alpha$ and $R$ when creating
a RGG. The distance between every pair of vertices must be
calculated, and an edge is added if the distance is less than $R$.
Thus, it seems unavoidable to have a runtime of $O(N^2)$ making
it unfeasible to investigate as large systems as with random
graphs---see Fig.~\ref{RGlargestc}---where the
number of calculations for a given $\alpha$ needed to create
all the edges is
$O(N)$. To overcome this obstacle we have designed a data
structure which is described in 
Section~\ref{sec:Implementation},
with a runtime of $O(N^{\beta})$ where $\beta\simeq1.3$. 
This allows us to study RGGs with up to $N=4^{11}>4\cdot10^6$ 
vertices, which
is more than an order of magnitude larger than usually 
accomplished~\cite{rintoul97}.

\section{Results} \label{sec:Results}
%
In our simulations of RGGs we def\mbox{}ine  $\alpha_c$ to be the 
lowest connectivity at which the fraction of vertices in 
the largest cluster is $>0$ in the macroscopic limit.
We make the bold claim that the systems we are able to analyse
consist of enough points to make  the critical connectivity
almost as sharply def\mbox{}ined as in 
Fig.~\ref{RGlargestc}. However, our main purpose is not to derive
high precision percolation thresholds. Instead, we are
more interested in the critical connectivity as a
function of the dimension of the RGGs.

In this paper
we express our threshold values in terms of $\alpha$.
Other popular choices are
the fractional volume $s$ occupied by the spheres~\cite{balberg87} 
or the density $N$ of spheres. 
The relation between 
these parameters 
at the percolation threshold is 
\begin{equation}
	\alpha_c=N_cV_{ex}=-2^d\ln(1-s_c) 
\label{frac_volume}
\end{equation}
(see e.g.\ \cite{xia88} for a derivation).
Usually, in continuum percolation 
the volume $V$ of each sphere
is f\mbox{}ixed while $N$ is the independent variable in a system
of size $[0,L]^d$.
The approach of measuring $N_c$ or $s_c$ 
for various values of $L$ has been used in both
two~\cite{gawlinski81} and three~\cite{rintoul97} 
dimensions, i.e.\ for discs and spheres, where the critical values  
are determined by the use of f\mbox{}inite size scaling. This procedure
resembles site percolation in lattices.
From the previous sections it is clear that we take 
a route closer to bond percolation in lattices
by f\mbox{}ixing $L=1$ while tuning
\al\ for dif\mbox{}ferent values of $N$.
In Section~\ref{sec:Implementation} we describe how this
has been carried out in practice. 
 
\begin{figure}[t]
\begin{center}
\includegraphics[width=7.5cm]{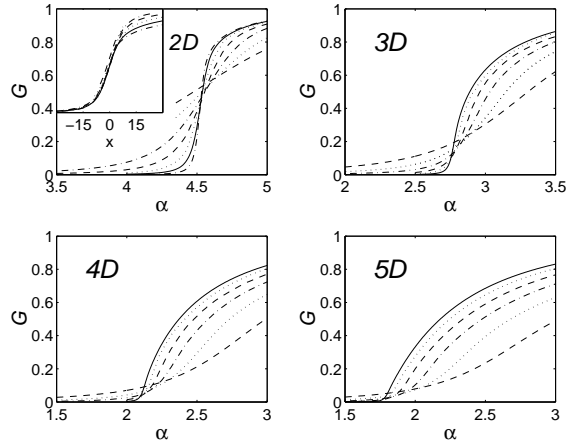}
\caption{\small 
	The average fraction of vertices in the	largest cluster
	for various system sizes $N$ (see the legend in
	Fig.~\ref{avlargestc2345dt}) in 
	random geometric graphs with no edges across the boundaries.
	The inset in $2D$ illustrates a f\mbox{}inite size 
	scaling---see the text.
	In higher dimensions the general shape
	of the curves as $N$ increases is nontrivial.
	Compare with Fig.~\ref{avlargestc2345dt}. 
	Error bars are $<10^{-3}$ for all curves and therefore omitted.
	}
\label{avlargestc234d}
\end{center}
\vspace{-0.5cm}
\end{figure}

\subsection*{The Size of the Largest Cluster}
\begin{figure}[t]
\begin{center}
\includegraphics[width=7.5cm]{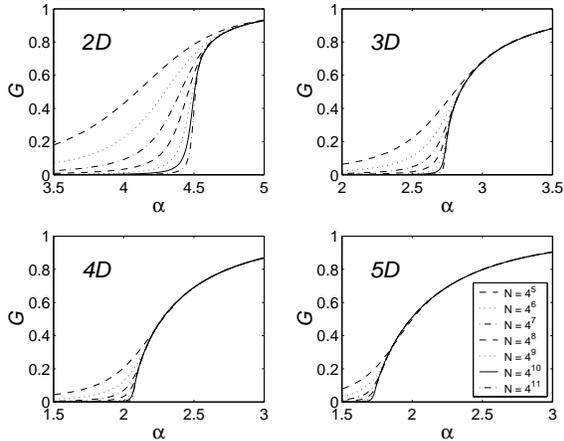}
\caption{\small 
	Like Fig.~\ref{avlargestc234d} but with \emph{continuous}
	boundary conditions. We see that the point at which
	the largest cluster becomes macroscopic is sharply def\mbox{}ined
	and can immediately be determined by the eye with high
	precision (Table~\ref{table_alpha_c}).
	The overall behaviour of the graphs for
	higher dimensions is much closer to Fig.~\ref{RGlargestc}
	than Fig.~\ref{avlargestc234d} is. As $d$ increases the
	$\alpha$-interval where there is a signif\mbox{}icant
	dif\mbox{}ference between curves with dif\mbox{}ferent $N$ 
	get smaller and smaller.
	Error bars are $<10^{-3}$ for all curves and therefore omitted.
	}
\label{avlargestc2345dt}
\end{center}
\vspace{-0.5cm}
\end{figure}

Let $G_d(\alpha)$ denote the fraction of vertices in the 
largest cluster in $d$ dimensions.
Since a RGG in the limit 
of inf\mbox{}inite dimension is often assumed 
equivalent to a random graph, we expect that
Eq.~(\ref{alpha(G)}) provides us with 
an expression for $G_{\infty}(\alpha)$.
But what does $G_d(\alpha)$ look like for f\mbox{}inite $d$?
And what is the behaviour of $\alpha_{c}(d)$? How does
it approach $\alpha_{c}(\infty)$ as $d$ increases? 
These are the questions addressed in 
this and the following section.

Figs.~\ref{avlargestc234d} and~\ref{avlargestc2345dt} illustrate
the average size of the largest cluster
in RGGs in 2, 3, 4, and 5 dimensions with and without 
toroidal boundary conditions. 
The curves correspond to $N=4^k$ vertices with $k=5,6,...,11$,
 where the larger systems display the sharpest transitions. The legend
in Fig.~\ref{avlargestc2345dt} applies to all diagrams in 
Figs.~\ref{avlargestc234d} and~\ref{avlargestc2345dt}. 
In these 8 diagrams each curve is
based on 300 data points. In other words, $G_d(\alpha)$ is 
calculated in intervals of $\Delta\alpha = 0.005$ resulting in the
smooth lines in the f\mbox{}igures. For every data set we have averaged
over enough runs for error bars to be completely negligible.

Since continuous boundary conditions mean addition of extra edges,
the size of the largest component $G(\alpha)$ obviously grows faster 
in Fig.~\ref{avlargestc2345dt} than in Fig.~\ref{avlargestc234d}, 
especially in the smaller systems. 
These relatively few extra edges make a decisive dif\mbox{}ference,
connecting vertices not already in the same cluster. Since toroidal
systems are models of bulk systems,
$G$ is much less $N$-dependent in that case.
However `unphysical' RGGs with open boundaries may seem, they
are the most popular RGG version in the literature.
Consequently, we consider them alongside the continuous case.

From Figs.~\ref{avlargestc234d} and~\ref{avlargestc2345dt} we see
that the continuous boundary conditions make the transition 
where $G>0$ more abrupt, but that an estimation of $\alpha_c$ does not 
depend much on the boundary conditions if only we base our judgment
on large enough systems.
This is conf\mbox{}irmed in the inset of Fig.~\ref{avlargestc234d}, where
$\alpha_c=4.53$ is obtained by f\mbox{}inite size scaling, 
i.e.\ plotting $G(x)$ where $x = N^{1/\nu}(\alpha-\alpha_c)$. 
However, it is clearly easier to make precise estimates of the 
critical connectivity with than without 
continuous boundary conditions.
We note in passing
that the exponent $\nu=3$ is equal to the value of $\nu$ found in
random graphs~\cite{kirkpatrick94}.

\subsection*{The Critical Connectivity} \label{subsec:scaling}

\begin{figure}[t]
\begin{center}
\includegraphics[width=7.5cm]{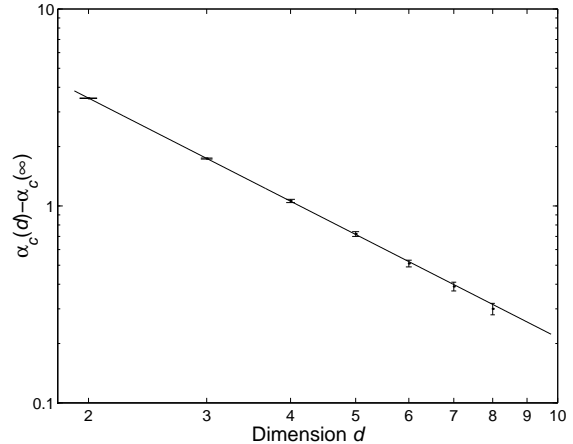}
\caption{\small 
	Scaling of the critical connectivity as a function of
	the dimension of the random geometric graphs
	reveals a power-law relation, Eq.~(\ref{alfa(d)}).
 	For $d\leq5$ the data points are estimated by
	close inspection of Fig.~\ref{avlargestc2345dt}.
	For $d>5$, \alc\ is based on runs with $N=4^{10}$ points.
	Error bars are included. See Table~\ref{table_alpha_c}.
	}
\label{alfa_c}
\end{center}
\vspace{-0.5cm}
\end{figure}

\noindent With numerically obtained knowledge of $G(\alpha)$, 
it is possible to extract $\alpha_c$. 
%
The procedure is simple. By inspection of Fig.~\ref{avlargestc2345dt}
we can estimate  \alc\ for $d\leq5$. To obtain further 
data points we have run our algorithm on  RGGs with $N=4^{10}$ for
systems of larger dimensions as well. Though this
results in increased runtime per graph, the results get more 
homogeneous and fewer runs are needed in order
to get a decent estimate of $G_d(\alpha)$.
Our f\mbox{}indings presented in Table~\ref{table_alpha_c}  
and Fig.~\ref{alfa_c} strongly suggest that
\begin{equation}
	\alpha_c(d) = \alpha_c(\infty) + Ad^{-\gamma},
\label{alfa(d)}
\end{equation}

\noindent where $\alc(\infty)=1$, 
$\gamma=1.74(2)$ and $A=11.78(5)$.
As expected, Eq.~(\ref{alfa(d)}) predicts that $\alc(\infty)$ 
is equal to $\alc$ in
random graphs, conf\mbox{}irming that RGGs and random graphs
become more and more similar as $d$ increases. However, 
when we derive the cluster coef\mbox{}f\mbox{}icient, 
we will see that this is not true in all respects.

Finally, we note that our f\mbox{}indings are 
in accordance with 
%
the most precise estimates 
that we know of: $\alpha_c=4.51223(5)$~\cite{quintanilla00} 
and  $\alpha_c=2.734(6)$~\cite{rintoul97}
in $2D$ and $3D$, respectively, obtained by
the use of f\mbox{}inite size scaling. 
For $d>3$ we have not been able to f\mbox{}ind any 
estimates of $\alpha_c$ to compare with~\cite{torquato02}.

\subsection*{The Distribution of Cluster Sizes}
\begin{table}
\begin{tabular}[t]{||l |l |l |l |l |l |l |l ||r}	\hline
$d$	 	&  2 & 3 & 4 & 5 & 6 & 7 & 8\\	\hline
$\alpha_c$  	&  4.52 & 2.74 & 2.06 & 1.72 & 1.51 & 1.39 & 1.30\\ \hline
$\pm$		&  0.01 & 0.01 & 0.02 & 0.02 & 0.02 & 0.02 & 0.02\\ \hline 
\end{tabular}
\caption{\small{
	The critical connectivity \alc\ in random geometric graphs
	of dimension $d$ with continuous 
	boundary conditions. The data are
	plotted in Fig.~\ref{alfa_c}. The estimated errors 
	in $\alc$ in the last row are rather conservative.
	}		
	}
\label{table_alpha_c}
\end{table}
Having examined the size of the largest cluster and the
critical connectivity, we now look at the distribution
of cluster sizes in RGGs.

\begin{figure}[t]
\begin{center}
\includegraphics[width=7.5cm]{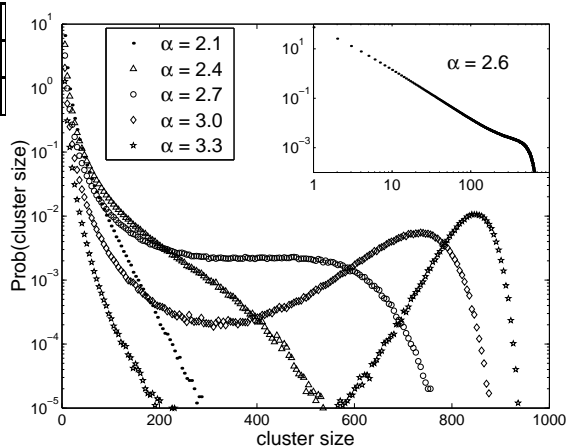}
\caption{\small 
	The distribution of cluster sizes in $3D$ 
	random geometric graphs with $N=1000$ vertices
	in the vicinity of the critical connectivity $ \alpha_c = 2.74$.
	The inset shows that for $\alpha \simeq \alpha_c$
	the cluster sizes are given by a power-law.
	For each value of $\alpha$ the data points are based
	on $10^6$ graphs.
	}
\label{cdist}
\end{center}
\vspace{-0.5cm}
\end{figure}

The inset illustrates the scale free 
power-law distribution at $\alpha=2.6$. Right below $\alpha_c$,
clusters of all sizes can be encountered. The small hump at large
cluster sizes is always present because the clusters cannot contain
more than all of the vertices. The clusters pile up when 
their size approaches this boundary, in this case a cluster size of
$1000$, just below the inevitable cut-of\mbox{}f.

Our simulations show that for $\alpha$ signif\mbox{}icantly below
$\alpha_c$ the distribution is approximately exponential.
As the connectivity increases the distribution 
becomes power-law-like. As $\alpha$ is further
increased the distribution is separated in two parts; 
there are no clusters of medium size, only the largest
macroscopic cluster and a few small ones around it. 
We have observed this overall behaviour in all our tests of the
distribution of cluster sizes in various dimensions. 

Fig.~\ref{cdist} shows our data in $3D$.
For $\alpha=2.1$ ($\cdot$) the data points lie on an almost straight
line indicating an exponential distribution. Increasing the
connectivity to $\alpha=2.4$ ($\triangle$)
results in a broader distribution that
is no longer  exponential. Right at the critical connectivity ($\circ$)
the distribution f\mbox{}lattens out. Clusters of all sizes are observed.
Right above $\alpha_c$ ($\diamond$) two separate regions 
begin to materialise. Already at
$\alpha=3.3$ ($\star$) the largest cluster makes it highly unlikely that
a cluster of medium size can be present as well. The distribution is
cut in two.

\subsection*{The Cluster Coef\mbox{}f\mbox{}icient}
In network theory
the cluster coef\mbox{}f\mbox{}icient $C$ is an often calculated
quantity~\cite{barabasi01,watts99,newman00c},
which is 
def\mbox{}ined in the following way. Let the vertices
$i$ and $j$ be connected directly to a common vertex $k$. $C$ is then
the probability that vertex $i$ and vertex $j$ 
are directly connected as well. 
From this we see that the cluster coef\mbox{}f\mbox{}icient is a measure of
the `cliquishness' of the graph. In this section we
derive $C=C_d$ analytically in arbitrary dimensions $d$, showing that
$C_d$ decreases in an exponential fashion.

To determine $C_d$ we make use of the
concept of the excluded volume $V_{ex}$. If we again
use the vertices $i$, $j$, and $k$, then $i$ and $j$ must both 
be within the excluded volume of $k$.
Put dif\mbox{}ferently, the probability that $i$ and $j$ are connected 
is equal to the probability that two randomly
chosen points in a sphere of volume
$V_{ex}$ and radius $R$ is less than a distance
$R$ apart. 
In other words, 
given the coordinates of vertex $i$
the probability that there is an edge between $i$ and $j$ is equal to
the fraction of the excluded volume of vertex $i$ that lies
inside the excluded volume of $k$. By averaging
over all points in $V_{ex}$ 
we get the cluster coef\mbox{}f\mbox{}icient $C_d$.

\begin{figure}[t]
\begin{center}
\includegraphics[width=7.5cm]{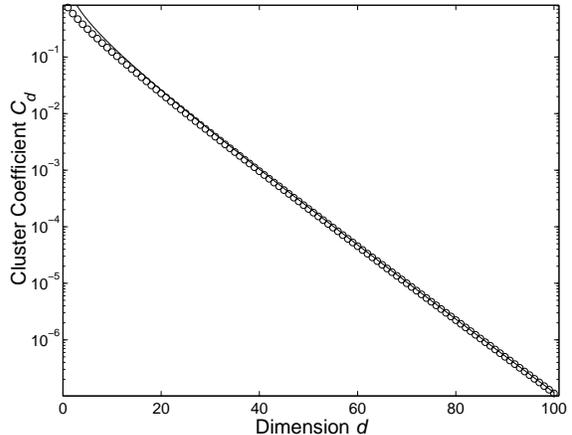}
\caption{\small 
	The cluster coef\mbox{}f\mbox{}icient $C$ 
	in random geometric graphs.
	The full line is the asymptotic solution,
	Eq.~(\ref{C_infty}), valid for large $d$ only.
	}
\label{clustcoeff}
\end{center}
\vspace{-0.5cm}
\end{figure}

The task of calculating $C_d$ is considerably simplif\mbox{}ied by
the spherical symmetry of the problem. 
The fractional volume `overlap' $\rho_d$
of two spheres only depends on the distance $r$ between the centers
and not on any angular parts, i.e.\ $\rho_d=\rho_d(r)$.
In general, the cluster 
coef\mbox{}f\mbox{}icient can therefore be written as
\begin{equation}
C_d 	= \frac{1}{V_{ex}} \int_{V_{ex}} \rho_d(r)\textrm{d}V.
\label{C_d}
\end{equation}
In Appendix~\ref{AppendixA} we derive that 
\begin{equation}
C_d = \left\{ 
	\begin{array}{ll}
	1 - H_d(1) & \mbox{even $d$} \\
	\\
	\frac{3}{2} - H_d(\frac{1}{2}) & \mbox{odd $d$}
	\end{array}
	\right.
\label{finalC_d}
\end{equation}
where
\begin{equation}
H_d(x) =\frac{1}{\sqrt{\pi}} \sum_{i=x}^{d/2} 
	\frac{\Gamma(i)}{\Gamma(i+\frac{1}{2})}
	\left( \frac{3}{4} \right)^{i+\frac{1}{2}}.
\label{H(x)}
\end{equation}
When $d$ is large Eq.~(\ref{finalC_d}) reduces to
(see Appendix \ref{AppendixA})
\begin{equation}
C_d \sim 3 \sqrt{\frac{2}{\pi d}} 
\left(\frac{3}{4}\right)^{\frac{d+1}{2}}.
\label{C_infty}
\end{equation}

\noindent The cluster coef\mbox{}f\mbox{}icient
is plotted in Fig.~\ref{clustcoeff} ($\circ$)
together with the asymptotic solution 
in Eq.~(\ref{C_infty}) (full line).

Eq.~(\ref{finalC_d}) shows that the cluster coef\mbox{}f\mbox{}icient
is a purely geometric quantity depending only on the dimension $d$;
neither the connectivity $\alpha$ nor the system size $N$ are
present. In random graphs $C=\alpha/N$, since there
is per def\mbox{}inition no correlation between edges.
So, in contrast to what is usually believed,
RGGs  are \emph{not} identical
to random graphs when $d \rightarrow \infty$.

In higher dimensions, the cluster 
coef\mbox{}f\mbox{}icient in RGGs becomes exceedingly small.
This peculiar fact can be explained by noting that 
the distribution of distances between 
two connected vertices gets more and more peaked at the 
maximal distance $R$ as $d$ increases. 
This implies that if the vertices $i$ and $j$ are both
connected to vertex $k$ in a high-dimensional space,
then it is highly unlikely that $i$ and $j$ are directly
connected by an edge as well.
Only in low dimensions are RGGs dominated by small loops.
On the contrary, 
the way that a standard random graph is designed
implies a cluster coef\mbox{}f\mbox{}icient which can only be 
interpreted statistically, and not geometrically.
Despite the fact that 
$\alpha_c=1$ in both random graphs 
and RGGs of infinite dimensionality, 
they do not have the same topology.

\subsection*{Graph Bi-partitioning} \label{sec:GBP}
Random geometric graphs are useful outside network modeling and 
percolation theory as well. In this section we look at RGGs in relation
to graph bi-partitioning, a well known problem in combinatorial
optimization.

The NP-hard problem of partitioning a graph with $N$ vertices in two
subsets with $N/2$ vertices each, 
in such a way that the cutsize $E$,
i.e.\ the number of edges between vertices in dif\mbox{}ferent 
subsets, is minimized, 
is called the graph bi-partitioning (GBP) problem.
Fig.~\ref{ggplot5} illustrates a bi-partitioned RGG, where
$N/2$ of the points are marked by squares, the other half being dots.

The GBP problem of RGGs
with open boundary conditions has been tested by various 
heuristics~\cite{johnson89,merz98,boettcherpercus00b}. In this
section we use our numerical 
f\mbox{}indings to establish the critical
connectivity in relation to GBP.
Additionally, for $\alpha > \alpha_c^{\scriptscriptstyle{GBP}}$
we argue that the cutsize $E$ depends on $N$ and $\alpha$ in
a simple way.

In GBP the connectivity is critical when $G=1/2$. As soon
as the largest cluster contains more than half of the vertices, it
becomes impossible to bipartition the graph without violating any edges.
For random graphs Eq.~(\ref{alpha(G)}) immediately gives us
$\alpha_c^{\scriptscriptstyle{GBP}}=2 \ln 2 \simeq 1.386$. 

\begin{table}
\begin{tabular}[t]{||l |l |l |l |l ||r}	\hline
$d$		 &  2 & 3 & 4 & 5 \\	\hline 
$\alpha_c^{\scriptscriptstyle{GBP}} $ 
		 &  4.52 & 2.84 & 2.275 & 1.99\\ \hline
$\pm$		 &  0.02 & 0.01 & 0.005 & 0.005  \\ \hline
\end{tabular}
\caption{\small{
	The critical connectivity $\alpha_c^{\scriptscriptstyle{GBP}}$
	in random geometric graphs with toroidal boundary conditions.
	Only in $2D$ does $\alpha_c^{\scriptscriptstyle{GBP}}$ 
	depend noticably on $N$ for $N>1000$ 
	(see Fig.~\ref{avlargestc2345dt}). 
	Note that without continuous boundaries
	Fig.~\ref{avlargestc234d} shows that 
	$\alpha_c^{\scriptscriptstyle{GBP}}$ is
	highly size-dependent for $d>2$.
	The estimated errors in $\alpha_c^{\scriptscriptstyle{GBP}}$ 
	in the last row are on the safe side.
	}		
	}
\label{table_GBP}
\end{table}

In RGGs $\alpha_c^{\scriptscriptstyle{GBP}}(d)$ 
can be extracted in the same way as 
$\alpha_c$ was in Section~\ref{subsec:scaling}. 
Our numerical f\mbox{}indings 
in RGGs with continuous boundary conditions are presented
in Table~\ref{table_GBP}. We stress that the results are
valid only for large $N$,
as a closer look
at Fig.~\ref{avlargestc2345dt} reveals. In $2D$ the average fraction
of vertices in the largest cluster is independent of $N$ only for 
$\alpha > \alpha_c^{\scriptscriptstyle{GBP}}$. 
This means that if one looks at GBP in $2D$
with $N=1000$, one cannot use the value of 
$\alpha_c^{\scriptscriptstyle{GBP}}$ in Table~\ref{table_GBP}. 
In higher dimensions  the interval
around $\alpha_c$ where $G_d(\alpha)$ is size-dependent gets smaller
and does not play a role in relation to GBP. 

With open boundary conditions the picture is messy, as 
Fig.~\ref{avlargestc234d} shows. In this case $G(\alpha)$ is highly
$N$-dependent, and it is not possible to speak of a critical
connectivity $\alpha_c^{\scriptscriptstyle{GBP}}$ without specifying $N$. 
This is true
despite the fact that $G(\alpha)$ is an averaged quantity, i.e.\ for
small $N$ will a fraction of the graphs 
contain a cluster with more than $N/2$
vertices even when $\alpha < \alpha_c^{\scriptscriptstyle{GBP}}$. 
Fig.~\ref{avlargestc234d} clearly shows that 
$\alpha_c^{\scriptscriptstyle{GBP}}$ is a decreasing function of $N$
for $d>2$. In $2D$ however,
all curves cross at almost the same (pivotal) point, and
it is reasonable to speak of $\alpha_c^{\scriptscriptstyle{GBP}}$ 
without specifying $N$. As the inset in Fig.~\ref{avlargestc234d}
shows this would lead to an estimate of 
$\alpha_c^{\scriptscriptstyle{GBP}}=4.53(1)$, 
close to $\alpha_c^{\scriptscriptstyle{GBP}}$ in RGGs with 
toroidal boundary conditions. 

The size of the largest cluster near $\alpha_c$ grows so rapidly 
in $2D$  that $\alpha_c=\alpha_c^{\scriptscriptstyle{GBP}}$
cannot be ruled out on the basis of our numerical data. This is
true with both open and continuous boundary conditions.
However, as this would 
imply that the phase transition is of 1st order in $2D$ only, we
believe that the two critical connectivities are close but not identical.

When bi-partitioning a RGG, it is obvious that the 
`area of contact'~\cite{boettcherpercus01b}
between the two subsets in the optimal conf\mbox{}iguration must be close
to a minimum. In $2D$ this means that the best achievable partition
must be close to simply cutting the graph in two at 
the coordinate values $x_1=1/2$ or $x_2=1/2$. 
This observation is especially 
relevant for large connectivities where the cutsize
is, f\mbox{}luctuations neglected, proportional 
to the length of the dividing 
line. All this tentatively indicates
how the cutsize $E$ in GBP behaves as a function of $N$ and $\alpha$
by looking at RGGs partitioned at $x_i=1/2$, where 
$1 \leq i \leq d$.
As we are about to argue, 
we expect a scaling relation like~\cite{boettcherpercus99a,boettcher99b}
\begin{equation}
	E_d \propto N^{1/\nu} \alpha^{\beta}(d),
\label{scale_GBP_Nalfa}
\end{equation}

\noindent where the exponents $\nu$ and $\beta$ only depend on the
dimension of the RGG.  

The exponents in Eq.~(\ref{scale_GBP_Nalfa})
can be determined in the following way.
Given the radius $R$ of the excluded volume of each vertex,
 the cutsize must
be proportional to $NR$, since only vertices with $1/2-R<x_i<1/2$
contribute to the cutsize 
(to avoid counting the violated edges twice we only look at the
vertices at one side of the partitioning plane at $x_i=1/2$), times
the average number of violated edges per vertex in this region,
which is proportional to $NR^d$. In other words,
\begin{equation}
	E_d \propto N^2 R^{d+1}.
\label{scale_GBP_NR}
\end{equation}

\noindent If instead of $R$ we want to express the
result in terms of $\alpha(d) \propto N R^d$, 
we get 
\begin{equation}
	1/\nu=1-\frac{1}{d}, \;\;\; \beta=1+\frac{1}{d}.
\label{mu_nu}  
\end{equation}

\noindent Since $E \propto N^2$ in Eq.~(\ref{scale_GBP_NR}),
the relation $1/\nu + \beta = 2$ holds in arbitrary dimensions.

Now, it is obvious that the scaling Ansatz is reasonable
only for $\alpha > \alc^{\scriptscriptstyle{GBP}}$. 
As Fig.~\ref{ggplot5} illustrates,
the optimal partition at $\alpha \sim \alc^{\scriptscriptstyle{GBP}}$ 
is highly complex and
not at all close to a straight line. If we incorporate that $E=0$
for $\alpha<\alc^{\scriptscriptstyle{GBP}}$ and 
replace Eq.~(\ref{scale_GBP_Nalfa}) with
\begin{equation}
E_d \propto N^{1/\nu} (\alpha(d)-\alc^{\scriptscriptstyle{GBP}})^{\beta},
\label{scale_GBP_SB}
\end{equation}
we do not expect
Eq.~(\ref{mu_nu}) to hold if we focus only on a region
near the critical connectivity. 
By the use of extremal
optimization, a heuristic that works particularly well near
phase transitions in hard combinatorial problems, 
Boettcher and Percus~\cite{boettcherpercus99a,boettcher99b} 
have found $\alpha_c^{\scriptscriptstyle{GBP}} \simeq 4.1$,  
$1/\nu \simeq 0.6$ and  $\beta \simeq 1.4$ in 
$2D$ for $4<\alpha<6$, 
not far of\mbox{}f our estimates in Eq.~(\ref{mu_nu}) 
valid for large connectivities. Note that the low estimate of 
$\alpha_c^{\scriptscriptstyle{GBP}}$ is expected; 
the algorithm does not always f\mbox{}ind the best partition,
and some graphs with $\alpha<\alc$ does have $E>0$.

\section{Implementation} \label{sec:Implementation}
The implementation is of major importance when studying
random geometric graphs, since a straightforward check of all
possible edges between the $N$ points will result in 
unfeasible runtimes $O(N^2)$. We now outline how our program works
and describe how to avoid runtimes $O(N^2)$.

The main idea is to divide and conquer.
Partition the $d$-dimensional box
in smaller subboxes and determine which
subbox each vertex belongs to. Given the connectivity and thereby the 
radius $R$ of the excluded volume, for each vertex
we then only have to look for potential edges to vertices 
in the subboxes adjacent to the subbox where
the vertex itself is located. This leads to a huge reduction in the
number of comparisons. And this just gets better when $N$ increases,
resulting in a decrease in $R$ as we saw in Fig.~\ref{critradius}.
By partitioning the box further as $N$ increases we avoid
a linear increase in the number of comparisons per vertex, which
would lead to the undesirable $O(N^2)$ growth.

The algorithm used when looking at RGGs is simple.
It works like this:
\begin{enumerate}
\item{Generate $d$ coordinates for each vertex.}
\item{Partition the space in small subboxes.}
\item{Find the edges.}
\item{Calculate the relevant quantities 
	($G$, cluster sizes etc.)\ as $\alpha$ increases.}
\end{enumerate}
\begin{figure}[t]
\begin{center}
\includegraphics[width=7.5cm]{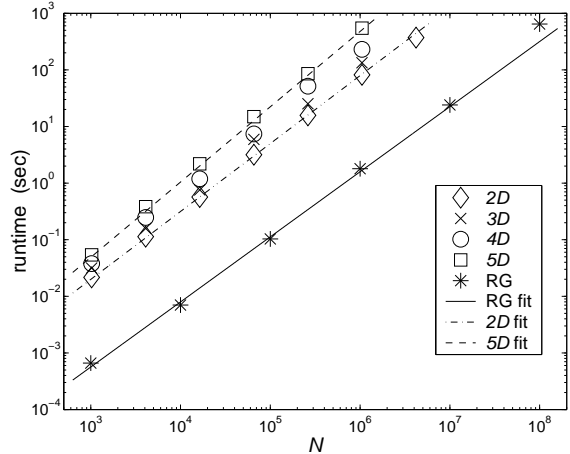}
\caption{\small 
	The runtimes (on a 400MHz SUN) 
	of the algorithms used in 
	Sections~\ref{sec:RG} and~\ref{sec:Results}.
	The straight lines indicate
	$t \sim N^{\beta}$,
	where $\beta=1.2$ in $2D$, $\beta=1.33$ in $5D$ and
	$\beta=1.15$ in random graphs (RG).
	}
\label{time}
\end{center}
\vspace{-0.5cm}
\end{figure}
Obviously, a trade-of\mbox{}f in Step 2 is involved 
when choosing the number of small boxes.

Being the most time consuming part of the algorithm, 
Step 3 is the main contributor when deciding 
how the runtime depends on $N$. The runtimes for most of
our runs are shown in Fig.~\ref{time}. We see that the runtime 
is $O(N^{\beta})$, where $\beta \approx 1.3$, resulting
in `feasible' runtimes for graphs with up to $N>4 \cdot 10^6$. Note
that the runtime of the much simpler algorithm used on random graphs 
also grows like a power-law with $\beta = 1.15$, even though the
number of operations is clearly $O(N)$. In fact, the number of
comparisons with potential neighbours per vertex is very
nearly constant in our implementation, i.e.\ 
the total number of neighbour tests is $O(N)$ in RGGs as well.
Of course, this is only possible if the number of
subboxes also increases with $N$. 
Managing the partitioning part of the algorithm adds to
the runtime. To sum up, the power-law increase in the
runtime illustrated in Fig.~\ref{time} for both random graphs
and RGGs is probably mainly due to cache misses.
The slightly higher values of $\beta$ in the RGGs stems from the
additional time used when partitioning 
the $d$-dimensional box into smaller boxes. 

Step 4 is worth a comment. When running the algorithm we are interested
in information at certain values of $\alpha$.
Instead of generating a new graph for every data point needed, 
we f\mbox{}irst set up the graph with the minimal connectivity we want to
look at.
This is
easily accomplished with our algorithm. Given an $\alpha$-window
$[\alpha_{min},\alpha_{max}]$ in which we want to examine the graph, 
we f\mbox{}ind all the edges belonging to the graph
when $\alpha=\alpha_{max}$, but we
only add the edges corresponding to $\alpha=\alpha_{min}$.
The rest of the edges, those who are
to be added when $\alpha$ is gradually
increased to $\alpha_{max}$, are stored in a priority queue. It is then
a simple task to increase $\alpha$ as one wishes. As mentioned earlier,
in Figs.~\ref{avlargestc234d} and~\ref{avlargestc2345dt} each curve
is based upon 300 data points, i.e.\ $\Delta \alpha=0.005$.

The source code, written in {\bf }C, is available upon request.
For a more accurate and
technical discussion of fast algorithms in relation
to RGGs, see e.g.~\cite{dickerson96}.

\section{Summary} \label{sec:Conclusion}
In this paper we have illustrated the usefulness of 
random geometric graphs in network theory and how 
to implement them ef\mbox{}f\mbox{}iciently.
Several properties of random geometric graphs in the
vicinity of the critical connectivity $\alpha_c$
have been analysed.
We have determined the size of the largest cluster numerically and
shown that $\alpha_c(d)$ approaches
$\alpha_c(\infty)=1$ found in random graphs in a power-law fashion.
We have verified that the distribution of cluster sizes
is cut in two just when 
the connectivity becomes larger than $\alpha_c$. 
Interestingly, the derivation of the 
cluster coef\mbox{}f\mbox{}icient shows that, even 
in the limit of inf\mbox{}inite dimensionality $d$,
random geometric graphs are not identical to random graphs.

Random geometric graphs share properties with both 
lattice models and standard random graphs. Random
geometric graphs allow us to work with random
graphs with a local structure. In addition, it is 
straightforward to add `long' edges if one
wishes to simulate, e.g.,\ 
a small world network. With all this in mind,
we hope this paper will make random geometric 
graphs more widely used in network theory.

\appendix 
\section{Derivation of $C_d$} \label{AppendixA}
\begin{figure}[t]
\begin{center}
\includegraphics[width=5.6cm]{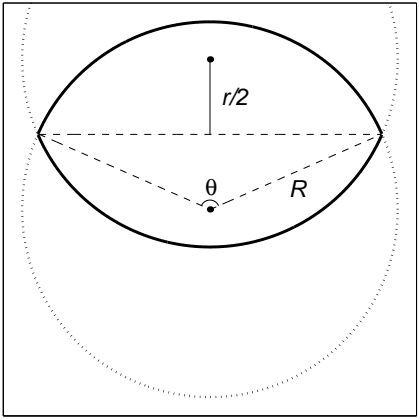}
\caption{\small 
	Determination of the cluster coef\mbox{}f\mbox{}icient $C$, which in
	$2D$ is equal to the average 
	fractional area overlap of the two circles. $R$ is
	the radius of the circles and $r$ the distance
	between their centers.
	The area of the overlap is  conf\mbox{}ined within
	the fat arcs originating from the two circles (dotted).
	The dashed lines are helpful in the derivation of
	the overlap---see the text.
	}
\label{overlap}
\end{center}
\vspace{-0.5cm}
\end{figure}

In order to determine the cluster coef\mbox{}f\mbox{}icient 
for arbitrary
$d$, one must find the fractional overlap $\rho_d$. 
Since $\rho_d$ has no angular dependence, Eq.~(\ref{C_d}) 
reduces to
\begin{equation}
C_d 	= \frac{d}{R^d} \int_0^R \rho_d(r)r^{d-1}\textrm{d}r. 
\label{C_dreduced}
\end{equation}
Since $\rho_1=1-\frac{r}{2R}$, $C_1=\frac{3}{4}$. 
From Fig.~\ref{overlap} we see that in $2D$ the overlapping
area---the area circumscribed by the fat lines---is 
$2(A-B)$, where $A$ is the area of the part
of the circle swept out by the angle
\mbox{$\theta = 2\arccos(r/2R)$} between
the two dashed lines originating
from the center of the lowest circle, and $B$ is the area
of the dashed triangle. Now, $A=\frac{1}{2}\theta R^2$ and 
$ B = R^2 \cos(\theta/2) \sin(\theta/2) = \frac{1}{2} R^2 \sin\theta $. 
The area of the overlap is then $ R^2(\theta - \sin\theta)$, so   
$\rho_2 = \frac{1}{\pi}(\theta - \sin\theta)$
and 
$C_2 = 
1\! -\! \frac{3\sqrt{3}}{4\pi}.$

For $d\geq3$, the use of cylindrical coordinates and 
the relation
\begin{equation}
2 \pi \prod_{i=2}^{n-1}
\int_{0}^{\pi} \sin^{n-i} \theta_i \textrm{d}\theta_i
=
\frac{n \pi^{n/2}}{\Gamma(\frac{n+2}{2})}
\label{aux1}
\end{equation}
results in
\begin{equation}
\rho_d(r) =
	\frac{2}{\sqrt{\pi}}
	\frac{\Gamma(\frac{d+2}{2})}{\Gamma(\frac{d+1}{2})}
	\int_{0}^{\arccos(\frac{r}{2R})} \sin^d \theta \textrm{d}\theta.
\label{rho_d>2}
\end{equation}
By reversing the integration in $C_d$ we get
\begin{equation}
C_d =
	\frac{3}{\sqrt{\pi}}
	\frac{\Gamma(\frac{d+2}{2})}{\Gamma(\frac{d+1}{2})}
	\int_{0}^{\frac{\pi}{3}} 
	\sin^d\theta \textrm{d}\theta,
\label{C_d>2}
\end{equation}
%
%
%
which can be solved by integration by parts.
The use of the duplicate formula for the Gamma function
then finally leads to Eq.~(\ref{finalC_d}).

For large $d$, the ratio of the Gamma functions 
in Eq.~(\ref{C_d>2})
is given by Stirling's approximation.
By putting $x=\cos \theta - 1/2$,
the cluster coef\mbox{}f\mbox{}icient can 
therefore be written as
\begin{equation}
C_d 	
\simeq
	\sqrt{\frac{6d}{\pi}} \!
	\left(
	\frac{3}{4}
	\right)^{\frac{d}{2}}
	\!\!
	\int_{0}^{\frac{1}{2}}
	\!
	\exp \left[ 
\frac{d\!-\!1}{2} 
	\ln f(x)
	\right]
	\! \textrm{d} x,
\label{aux3}
\end{equation}
where $f(x) = 1 - \frac{4x}{3}(1+x)$. Since the 
contributions to the integral for large $d$ are significant
only when $x \simeq 0$, $\ln f$ can be expanded to
1st order and Eq.~(\ref{C_infty}) is recovered.

\vspace{0.8cm}
\noindent {\Large{\textbf{Acknowledgements}}} \\
\vspace{0.02cm}

\noindent We thank J. Neil Bearden, 
Stefan Boettcher and Paolo Sibani for helpful comments.
We are especially indebted to Allon Percus for very valuable
discussions and correspondence.
This work was financially supported by 
Statens Naturvidenskabelige Forskningsr{\aa}d.

\vspace{0.2cm}

\bibliography{diverse,network,statmec,optimization,evolution,sg}

\begin{thebibliography}{10}

\bibitem{barabasi01}
Albert-L{\'{a}}szl{\'{o}} Barab{\'{a}}si and R{\'{e}}ka Albert.
\newblock Statistical mechanics of complex networks.
\newblock {\em cond-mat/0106096v1}, pages 1--54, 2001.

\bibitem{strogatz01}
Steven~H. Strogatz.
\newblock Exploring complex networks.
\newblock {\em Nature}, 410:268--276, 2001.

\bibitem{huberman99}
Bernardo~A. Huberman.
\newblock Growth dynamics of the {W}orld-{W}ide {W}eb.
\newblock {\em Nature}, 401:131, 1999.

\bibitem{newman01b}
M.~E.~J. Newman, S.H. Strogatz, and D.J. Watts.
\newblock Random graphs with arbitrary degree distributions and their
  applications.
\newblock {\em Physical Review E}, 64:026118, 2001.

\bibitem{barabasi99}
R{\'{e}}ka Albert and Albert-L{\'{a}}szl{\'{o}} Barab{\'{a}}si.
\newblock Emergence of scaling in random networks.
\newblock {\em Science}, 286:509--512, 1999.

\bibitem{newman01c}
M.~E.~J. Newman.
\newblock Clustering and preferential attachment in growing networks.
\newblock {\em Physical Review E}, 64:025102, 2001.

\bibitem{amaral00}
L.~A.~N. Amaral, A.~Scala, M.~Barth{\'e}l{\'e}my, and H.~E. Stanley.
\newblock Classes of small-world networks.
\newblock {\em Proc. Natl. Acad. Sci. USA}, 97:11149--11152, 2000.

\bibitem{watts98}
Duncan~J. Watts and Steven~H. Strogatz.
\newblock Collective dynamics of `small-world' networks.
\newblock {\em Nature}, 393:440--442, 1998.

\bibitem{kauffman95}
Stuart Kauffman.
\newblock {\em At Home in the Universe}.
\newblock Oxford University Press, 1995.

\bibitem{erdos60}
P.~Erd{\"{o}}s and A.~R{\'e}nyi.
\newblock On the evolution of random graphs.
\newblock {\em Publ. Math. Inst. Hung. Acad. Sci.}, 5:17--61, 1960.

\bibitem{newman01d}
Duncan~S. Callaway, John~E. Hopcroft, Jon~M. Kleinberg, M.~E.~J. Newman, and
  Steven~H. Strogatz.
\newblock Are randomly grown graphs really random?
\newblock {\em Physical Review E}, 64:041902, 2001.

\bibitem{bollobas85}
B{\'e}la Bollob{\'a}s.
\newblock {\em Random Graphs}.
\newblock Academic Press, 1985.

\bibitem{kirkpatrick94}
Scott Kirkpatrick and Bart Selman.
\newblock Critical behavior in the satisfiability of random boolean
  expressions.
\newblock {\em Science}, 264:1297--1301, 1994.

\bibitem{kleinberg00}
Jon~M. Kleinberg.
\newblock Navigation in a small world.
\newblock {\em Nature}, 406:845, 2000.

\bibitem{stauffer92}
D.~Stauffer and A.~Ahorony.
\newblock {\em Introduction to Percolation Theory}.
\newblock Taylor and Francis, 1992.

\bibitem{cohen00}
Reuven Cohen, Keren Erez, Daniel ben Avraham, and Shlomo Havlin.
\newblock Resilience of the {I}nternet to {R}andom {B}reakdowns.
\newblock {\em Physical Review Letters}, 85:4626--4628, 2000.

\bibitem{albert00}
R{\'{e}}ka Albert, Hawoong Jeong, and Albert-L{\'{a}}szl{\'{o}} Barab{\'{a}}si.
\newblock Error and attack tolerance of complex networks.
\newblock {\em Nature}, 406:378--382, 2000.

\bibitem{krapivsky00}
P.~L. Krapivsky, S.~Redner, and F.~Leyvraz.
\newblock Connectivity of {G}rowing {R}andom {N}etworks.
\newblock {\em Physical Review Letters}, 85:4629--4632, 2000.

\bibitem{albertbarabasi00}
R{\'{e}}ka Albert and Albert-L{\'{a}}szl{\'{o}} Barab{\'{a}}si.
\newblock Topology of {E}volving {N}etworks: {L}ocal {E}vents and
  {U}niversality.
\newblock {\em Physical Review Letters}, 85:5234--5237, 2000.

\bibitem{dorogovtsev00}
S.~N. Dorogovtsev and J.~F.~F. Mendes.
\newblock Evolution of networks with aging of sites.
\newblock {\em Physical Review E}, 62:1842--1845, 2000.

\bibitem{watts99}
D.~J. Watts.
\newblock {\em Small Worlds}.
\newblock Princeton University Press, 1999.

\bibitem{newman99}
M.~E.~J. Newman and D.~J. Watts.
\newblock Scaling and percolation in the small-world network model.
\newblock {\em Physical Review E}, 60:7332--7342, 1999.

\bibitem{newman00c}
M.~E.~J. Newman.
\newblock Models of the small world.
\newblock {\em Journal of Statistical Physics}, 101:819--841, 2000.

\bibitem{banavar99}
Jayanth~R. Banavar, Amos Maritan, and Andrea Rinaldo.
\newblock Size and form in efficient transportation networks.
\newblock {\em Nature}, 399:130--132, 1999.

\bibitem{xia88}
W.~Xia and M.~F. Thorpe.
\newblock Percolation properties of random ellipses.
\newblock {\em Physical Review A}, 38(5):2650--2656, 1988.

\bibitem{balberg85}
I.~Balberg.
\newblock {"U}niversal{"} percolation-threshold limits in the continuum.
\newblock {\em Physical Review B}, 31:4053--4055, 1985.

\bibitem{alonbalberg90}
U.~Alon, A.~Drory, and I.~Balberg.
\newblock Systematic derivation of percolation thresholds in continuum systems.
\newblock {\em Physical Review A}, 42:4634--4638, 1990.

\bibitem{alonbalberg91}
U.~Alon, I.~Balberg, and A.~Drory.
\newblock New, heuristic, percolation criterion for continuum systems.
\newblock {\em Physical Review Letters}, 66:2879--2882, 1991.

\bibitem{quintanilla00}
J.~Quantanilla, S.~Torquato, and R.~M. Ziff.
\newblock Efficient measurements of the percolation threshold for fully
  penetrable disks.
\newblock {\em Journal of Physics A: Math. Gen.}, 33:L399--L407, 2000.

\bibitem{balberg87}
I.~Balberg.
\newblock Recent developments in continuum percolation.
\newblock {\em Phil. Mag. B}, 56(6):991--1003, 1987.

\bibitem{jund01}
Philippe Jund, R{\'{e}}mi Jullien, and Ian Campbell.
\newblock Random walks on fractals and stretched exponential relaxation.
\newblock {\em Physical Review E}, 63:036131, 2001.

\bibitem{vespignani01c}
Romualdo Pastor-Satorras and Alessandro Vespignani.
\newblock Epidemic {S}preading in {S}cale-{F}ree {N}etworks.
\newblock {\em Physical Review Letters}, 86:3200--3203, 2001.

\bibitem{cheeseman91}
Peter Cheeseman, Bob Kanefsky, and William~M. Taylor.
\newblock Where the \textit{Really} {H}ard {P}roblems {A}re.
\newblock {\em Proc. of IJCAI-91}, pages 331--337, 1991.

\bibitem{anderson99}
Philip~W. Anderson.
\newblock Solving problems in finite time.
\newblock {\em Nature}, 400:115--116, 1999.

\bibitem{wong87}
K.~Y.~M. Wong and D.~Sherringham.
\newblock Graph bipartitioning and spin glasses on a random network of fixed
  finite valence.
\newblock {\em Journal of Physics A: Math. Gen.}, 20:L793--L799, 1987.

\bibitem{molloyreed95}
Michael Molloy and Bruce Reed.
\newblock A {C}ritical {P}oint for {R}andom {G}raphs with a {G}iven {D}egree
  {S}equence.
\newblock {\em Random Structures and Algorithms}, 6:161--179, 1995.

\bibitem{molloyreed98}
Michael Molloy and Bruce Reed.
\newblock The {S}ize of the {G}iant {C}omponent of a {R}andom {G}raph with a
  {G}iven {D}egree {S}equence.
\newblock {\em Combinatorics, Probability and Computing}, 7:295--305, 1998.

\bibitem{rintoul97}
M.~D. Rintoul and S.~Torquato.
\newblock Precise determination of the critical threshold and exponents in a
  three-dimensional continuum percolation model.
\newblock {\em Journal of Physics A: Math. Gen.}, 30:L585--L592, 1997.

\bibitem{gawlinski81}
Edward~T Gawlinski and H~Eugene Stanley.
\newblock Continuum percolation in two dimensions: {M}onte {C}arlo tests of
  scaling and universality for non-interacting discs.
\newblock {\em Journal of Physics A: Math. Gen.}, 14:L291--L299, 1981.

\bibitem{torquato02}
Salvatore Torquato.
\newblock {\em Random H{}eterogenouos {M}aterials: {M}icrostructure and
  {M}acroscopic {P}roperties}.
\newblock Springer, 2002.

\bibitem{johnson89}
David S.~Johnson \emph{et.\ al.}
\newblock Optimization by simulated annealing: An experimental evaluation; part
  1, graph bi-partitioning.
\newblock {\em Operations Research}, 37:865--892, 1989.

\bibitem{merz98}
Peter Merz and Bernd Freisleben.
\newblock Memetic algorithms and the fitness landscape of the graph
  bi-partitioning problem.
\newblock {\em Lecture Notes in Computer Science}, 1498:765--774, 1998.

\bibitem{boettcherpercus00b}
Stefan Boettcher and Allon~G. Percus.
\newblock Nature's way of optimizing.
\newblock {\em Artificial Intelligence}, 64:275--286, 2000.

\bibitem{boettcherpercus01b}
Stefan Boettcher and Allon~G. Percus.
\newblock Extremal optimization for graph partitioning.
\newblock {\em Physical Review E}, 64:026114, 2001.

\bibitem{boettcherpercus99a}
Stefan Boettcher and Allon~G. Percus.
\newblock Extremal optimization: Methods derived from co-evolution.
\newblock {\em GECCO-99: Proceedings of the Genetic and Evolutionary
  Computation Conference}, pages 825--832, 1999.

\bibitem{boettcher99b}
Stefan Boettcher.
\newblock Extremal optimization of graph partitioning at the percolation
  threshold.
\newblock {\em Journal of Physics A: Math. Gen.}, 32:5201--5211, 1999.

\bibitem{dickerson96}
Matthew~T. Dickerson and David Eppstein.
\newblock Algorithms for {P}roximity {P}roblems in {H}igher {D}imensions.
\newblock {\em Comp. Geom. Theory and Appl.}, 5:277--291, 1996.

\end{thebibliography}
\bibliographystyle{unsrt}

\end{document}